\documentclass[pra,twocolumn]{revtex4}

\usepackage{graphicx,epsfig,amssymb,multirow,amsthm,longtable}

\def\duzomniejsze{<\kern-.7mm<}
\def\duzowieksze{>\kern-.7mm>}

\def\textbf#1{{\bf #1}}
\def\be{\begin{equation}}
\def\ee{\end{equation}}
\def\ben{\begin{eqnarray}}
\def\een{\end{eqnarray}}
\def\eea{\end{array}}
\def\bea{\begin{array}}
\newcommand{\bei}{\begin{itemize}}
\newcommand{\eei}{\end{itemize}}
\newcommand{\bee}{\begin{enumerate}}
\newcommand{\eee}{\end{enumerate}}

\newtheorem{definition}{Definition}
\newtheorem{theorem}{Theorem}
\newtheorem{lemma}{Lemma}

\newtheorem{corollary}{Corollary}
\newtheorem{our_rule}{Rule}

\def\blacksquare{\vrule height 4pt width 3pt depth2pt}

\def\tcal{{\cal T}}

\def\dt#1{{{\kern -.0mm\rm d}}#1\,}

\def\tr{{\rm Tr}}

\def\>{\rangle}
\def\<{\langle}
\def\blacksquare{\vrule height 4pt width 3pt depth2pt}

\def\ot{\otimes}

\def\flip{flipping}

\newcommand{\lar}{\rotatebox[origin=c]{0}{$\circlearrowright$}}

\begin{document}

\title{Characterization of combinatorially independent permutation
separability criteria}

\author{Pawe\l{} Wocjan$^{1}$ and Micha\l{} Horodecki$^{2}$}

\affiliation{${}^1$ Computer Science Department \& Institute for
Quantum Information, California Institute of Technology, Pasadena, CA
91125, USA\\
${}^2$ Institute of Theoretical Physics and Astrophysics,
University of Gda\'nsk, 80--952 Gda\'nsk, Poland}

\begin{abstract}
The so-called permutation separability criteria are simple operational
conditions that are necessary for separability of mixed states of
multipartite systems: (1) permute the indices of the density matrix
and (2) check if the trace norm of at least one of the resulting
operators is greater than one.  If it is greater than one then the
state is necessarily entangled. A shortcoming of the permutation
separability criteria is that many permutations give rise to dependent
separability criteria. Therefore, we introduce a necessary condition
for two permutations to yield independent criteria called
combinatorial independence. This condition basically means that the
map corresponding to one permutation cannot be obtained by
concatenating the map corresponding to the second permutation with a
norm-preserving map. We characterize completely combinatorially
independent criteria, and determine simple permutations that represent
all independent criteria. The representatives can be visualized by
means of a simple graphical notation. They are composed of three basic
operations: partial transpose, and two types of so-called
reshufflings. In particular, for a four-partite system all criteria
except one are composed of partial transpose and only one type of
reshuffling; the exceptional one requires the second type of
reshuffling. Furthermore, we show how to obtain efficiently for every
permutation a simple representative. This method allows to check
easily if two permutations are combinatorially equivalent or not.
\end{abstract}


\maketitle

\section{Introduction}
Entanglement theory has been actively developed for more than a
decade. We know a lot about entanglement of bipartite and multipartite
states, about how it can be manipulated and how it can be used as a
resource for performing certain tasks in quantum information
processing \cite{Alber2001}. However, so far one of the main problems
of entanglement theory has not been resolved: the design of
operational criteria allowing to detect whether a given state is
entangled or not. Since the first treatments of this subject
\cite{Werner1989,HH94-redun} a rich literature has been created (see
e.g.
\cite{QIC,Bruss2001-reflections,BadziagDHHH01-conc,MintertKB04-conc,GuehneL04-sep,DohertyPS04}).

For the purpose of this present paper, let us explicitly mention that
a simple necessary and sufficient condition exists for qubit-qubit and
qubit-qutrit systems, which is the partial transposition criterion
\cite{Peres96,sep1996}. Subsequently, in \cite{Chen2002-criterion} and
\cite{Rudolph2002-criterion} a new, equally simple criterion for
bipartite systems was introduced, called  or {\it
cross norm} or {\it realignment} criterion. Though weaker in the case of two qubits, it
turns out to be independent of the partial transpose in the case of
two qutrits and higher dimensional systems.  Then the multipartite
case has been treated in \cite{HHH02-permut}, where a family of
permutation criteria was introduced, which present a generalization of
both the partial transpose and the realignment
criteria \footnote{Another generalization for multipartite states was
introduced in \cite{ChenWu02-multisep}.}.

The basic idea underlying the permutation separability criteria is
that the operators that are obtained by permuting the indices of the
density matrix of any separable state still have trace norm not
greater than one. If the norm is greater than one than the state is
necessarily entangled.
An advantage of the permutation criteria is that they are very easy to
apply.  However, some inconvenience still remains, both for direct
applications, as well as for gaining theoretical insight. The reason
for this is that many different permutations can lead to the same
separability criterion. Indeed, suppose that a given permutation
$\sigma$ of indices does not change the norm of any operator (for
example, because it gives rise to a unitary operation).  Consider now
some other permutation $\sigma'$.  Clearly, a permutation $\sigma'
\sigma$ where we first apply $\sigma'$ and then $\sigma$, gives the
same criterion as $\sigma'$.  Many permutation criteria are equivalent
in this sense. We call this {\it combinatorial (in)dependence}.  For
bipartite systems one easily finds that as the only combinatorially
independent permutation criteria one can choose partial transpose and
realignment \cite{HHH02-permut}. For multipartite systems the problem
of determining independent criteria turns out to be more difficult.
The problem of deriving independent criteria was attempted in
\cite{Fan-permut} where combinatorially independent criteria for the
tripartite case were determined and a method of deciding whether two
criteria are equivalent was exhibited.

In this paper we solve the problem of characterizing combinatorially
independent permutation criteria completely. For an $r$-partite
system, we show that there are ${1\over 2} {2r \choose r}-1$
combinatorially independent criteria. We provide a complete set of
independent criteria, which are particularly simple. They can be built
by combining three basic operations: partial transpose (acting on only
one of the subsystems) and two kinds of so-called reshuffling (both
acting on two of the subsystems). We introduce a simple graphical
notation for the independent criteria and show how a general
permutation can be brought into this form. As a result we obtain a
simple and transparent method of verifying whether two permutations
lead to equivalent criteria. In this way we have transformed the
permutation criteria in a user-friendly tool, showing in a precise
manner the structure of all combinatorially independent permutation
separability criteria.

\section{Definition of permutation separability criteria}
In this section we recall the definition and the properties of the
permutation criteria for separability of multipartite mixed states. We
consider states acting on $r$ subsystems whose Hilbert spaces have all
the same dimension $d$. A general state $\rho$ on this joint Hilbert
${\cal H}$ space can be written as follows
\begin{equation}
\sum_{i_1,i_2, \ldots, i_{2r-1},i_{2r}} 
\rho_{i_1 i_2, i_3 i_4, \ldots, i_{2r-1} i_{2r}}
|i_1 i_3\ldots i_{2r-1}\> \<i_2 i_4 \ldots i_{2r}|
\end{equation}
where all indices run from $1$ to $d$, and the kets $|k_1 \ldots
k_r\>=|k_1\>\ot |k_2\>\ot \ldots \ot |k_r\>$ denote the standard basis
vectors in ${\cal H}$. Note that indices $i_j$ with odd subscripts
correspond to rows of $\rho$ and those with even subscripts to columns
of the density matrix.

Let $S_{2r}$ denote the group of permutations of the set $\{1,2,
\ldots, 2r\}$. We define for each permutation $\sigma\in S_{2r}$ a
corresponding map $\Lambda_{\sigma}$ on operators acting on ${\cal H}$
by setting
\begin{eqnarray}
&& \big[
\Lambda_\sigma(\rho)
\big]_{i_1 i_2, i_3i_4, \ldots, i_{2r-1} i_{2r}} \nonumber \\ 
& = & 
\rho_{i_{\sigma(1)} i_{\sigma(2)}, i_{\sigma(3)}i_{\sigma(4)}, \ldots, 
i_{\sigma(2r-1)} i_{\sigma(r)}}\,.
\end{eqnarray}
Note that the map $\sigma\mapsto \Lambda_\sigma$ is a homomorphism
from the symmetric group $S_{2r}$ into the group of invertible
operators acting on ${\cal H}$, i.e., we have
\begin{equation}
\Lambda_{\sigma_1\sigma_2}(\rho)=
\Lambda_{\sigma_2}(\Lambda_{\sigma_1}(\rho)))\,,
\label{eq:homo}
\end{equation}
where $\sigma_1\sigma_2$ denotes the composition of the permutations
$\sigma_1$ and $\sigma_2$. Here and throughout the paper we use the
convention that products of permutations are evaluated from left to
right (i.e. $\sigma_1$ acts first, then $\sigma_2$).

According to \cite{HHH02-permut} each permutation $\sigma\in S_{2r}$
of indices of $\rho$ gives rise to a corresponding {\it permutation
criterion} of separability, as follows. It turns out that for
separable states we have
\begin{equation}
\|\Lambda_\sigma(\rho)\|\leq 1
\end{equation}
for all permutation $\sigma\in S_{2r}$, where $\|A\|=\tr (A
A^\dagger)^{1/2}$ denotes the trace norm.  Thus the above inequality
constitutes a separability criterion. If the inequality is violated,
it implies that the state $\rho$ is entangled.

\section{Combinatorial independence: a necessary condition for the independence of the permutation separability criteria}
As already noted in the introduction, a major inconvenience in
applying and understanding the permutation separability criteria stems
from the fact that many permutations define trivial separability
criteria or criteria that are not independent.

Let us first consider the case of trivial criteria. To do that, we
introduce some notions. We call a map $\Lambda$ on operators acting on
${\cal H}$ {\em norm-preserving} if $\|\Lambda(\rho)\|=\|\rho\|$ for all
operators $\rho$ on ${\cal H}$. Similarly, we say that a permutation
$\sigma$ is norm-preserving if its corresponding map $\Lambda_\sigma$
is norm-preserving. Now, let $\sigma$ be a norm-preserving
permutation. In this case $\|\Lambda_\sigma(\rho)\|=1$ for all quantum
states $\rho$ and consequently $\sigma$ defines a trivial
criterion. Such a criterion is useless because it cannot detect any
entanglement.

For example, this can occur if the map $\Lambda_\sigma$ acts as
\begin{equation}
\Lambda_\sigma(\rho)= U_\sigma \rho V_\sigma
\label{eq:u1u2}
\end{equation}
on all operators $\rho$ on ${\cal H}$, where $U_\sigma,V_\sigma$ are
unitary transformations. This is because multiplication by unitary
operators from the left or right cannot change the trace norm.

Another example for a trivial criterion is the transposition of the
density matrix of the state $\rho$ (defined as the exchange of rows
and columns). We call it {\em global quantum transposition} (GQT) in
order to differentiate it from partial (quantum) transpositions and
transpositions \footnote{We mean by a transposition a permutation of
the form $(a\,,b)$ and by a quantum transposition the transpose of a
matrix}. It is readily verified that the permutation that gives rise
to GQT is the following:
\begin{equation}
\sigma=(1,\,2)(3,\,4) \cdots (2r-1,\,2r)\,. 
\end{equation}

Let us now consider the problem of permutations leading to criteria
that are not independent. Assume that two different permutations
$\sigma$ and $\tau$ define two maps $\Lambda_\sigma$ and
$\Lambda_\tau$ that are related by a norm-preserving map
$\Lambda$. More precisely, assume that we have
\begin{equation}
\Lambda_\sigma(\rho) = \Lambda(\Lambda_\tau(\rho))
\end{equation}
for all operators $\rho$. It is clear that the separability criteria
defined by $\sigma$ and $\tau$ are equivalent because
\[
\|\Lambda_\sigma(\rho)\|=
\|\Lambda(\Lambda_\tau(\rho))\|=
\|\Lambda_\tau(\rho)\|
\]
for all quantum states. All entangled states that can be detected by
$\sigma$ can also be detected by $\tau$ and vice versa.

Based on the above observations we define combinatorial independence
of permutations that is a necessary condition for two permutations to
yield independent separability criteria.

\begin{definition}[Combinatorial independence]
We say that a permutation $\sigma$ in $S_{2r}$ (and its corresponding
separability criterion) is trivial if the corresponding map
$\Lambda_\sigma$ is norm-preserving. We say that two permutations
$\sigma$ and $\tau$ in $S_{2r}$ (and their corresponding separability
criteria) are combinatorially independent if and only if there is no
norm-preserving map $\Lambda$ on operators acting on ${\cal H}$ such
that
\begin{equation}
\Lambda_\sigma(\rho) = \Lambda(\Lambda_\tau(\rho))
\end{equation}
for all operators $\rho$, i.e., the map $\Lambda_\sigma$ is the
composition of $\Lambda_\tau$ with $\Lambda$ (where we apply first
$\Lambda_\tau$ and then $\Lambda$).
\end{definition}

In the following we derive a group-theoretical explanation of the
combinatorial independence relation. The property in
eq.~(\ref{eq:homo}) that the map $\sigma\mapsto\Lambda_\sigma$ is a
homomorphism ensures that norm preserving permutations form a subgroup
of $S_{2r}$. This is because the composition of two norm-preserving
permutations gives rise to the composition of two maps that do not
change the trace norm. We call this group the {\em group of
norm-preserving permutations} and denote it by $\tcal$.

\begin{lemma}[Group of norm-preserving permutations]\label{lem:group}
The group $\tcal$ of norm-preserving permutations is generated by
transpositions exchanging only even points, by transpositions
exchanging only odd points, and by the permutation realizing the
global quantum transposition:
\begin{equation}\label{eq:generation}
\big<\,(2k,\,2l),\, (2k-1,\,2l-1),\, (1,\,2)(3,\,4) \cdots (2r-1,\,2r)
\,\big>\,,
\end{equation}
where $1\leq k<l\leq r$. The group $\tcal$ has the structure of a
so-called wreath product: it contains two symmetric groups $S_r$
acting on even and odd numbers, respectively, and the subgroup $S_2$
generated by the permutation $(1,\, 2)(3,\, 4)\ldots (2r-1,\, 2r)$
that switches between the two symmetric groups. The number of elements
of the group $\tcal$ is equal to
\begin{equation}
2\,r!\,r!
\end{equation} 
Here the factor $2$ accounts for the permutation $(1,\, 2)(3,\,
4)\cdots (2r-1,\, 2r)$, and both $r!$ factors for the permutations
groups on even and odd points.
\end{lemma}
{\bf Proof}. The proof consists of two steps. We determine the group
structure by showing (1) that the group generated in
eq.~(\ref{eq:generation}) is a subgroup of the group of
norm-preserving permutations and (2) that all permutations that are
not contained in this subgroup change the trace norm of some quantum
states. Our graphical notation is decisive for the proof the second
part. It is postponed to the next section where it follows readily as
a corollary from Theorem~\ref{thm:character}.

We have already seen that the permutation $(1,\, 2)(3,\, 4)\cdots
(2r-1,\, 2r)$ is norm-preserving. It remains to be shown that the odd
permutations $(2k-1,\, 2l-1)$ and the even permutations $(2k,\,2l)$
are norm-preserving. To see that we show that their corresponding maps
can be realized by multiplication with unitary operators $V_{kl}$ from
left and right, respectively. Let $V$ denote the swap operator acting
on ${\mathbb C}^d\otimes \mathbb{C}^d$, i.e.,
\begin{equation}
V=\sum_{ij=1}^d |ij\>\<ji|
\end{equation}
The swap operator has the property $V(|\psi\>\otimes
|\phi\>)=|\phi\>\otimes |\psi\>$ for any $|\phi\>,|\psi\>\in
\mathbb{C}^d$. The swap operators between any two subsystems of ${\cal
  H}$ are defined by embedding $V$ suitably, i.e., the operators
$V_{kl}$ act as the swap operator $V$ on the subsystems $k$ and $l$
and as the identity on the remaining subsystems. Let us define two
maps on operators on ${\cal H}$ with the help of the swap operators:
\begin{equation}
V^L_{kl}(\rho)=V_{kl} \rho,\quad V^R_{kl}(\rho)= \rho V_{kl}\,.
\end{equation}
Note that the above maps on operators come from transpositions:
\begin{equation}
V^R_{kl}=\Lambda_\tau,\quad\quad V^L_{kl}=\Lambda_{\tau'} 
\end{equation} 
where $\tau=(2k-1,\,2l-1)$ and $\tau'=(2k,\,2l)$. For this reason the
transpositions exchanging two even or two odd numbers are
norm-preserving and therefore constitute trivial separability
criteria. Since compositions of norm-preserving permutations are again
norm-preserving permutations we see that all permutations of odd
numbers (and even numbers) are trivial permutations (this is because
transpositions generate all permutations). The wreath product
structure follows from the fact that if we multiply the permutations
that generate $\tcal$ we can always propagate the permutations
realizing the GQT to the right. Thus any element of the group can be
see as either a composition of some permutation of even indices and
some permutation of odd indices, or such a composition followed by
GQT.  Thus the number of elements is $2r!r!$. \quad\blacksquare

\begin{lemma}[Group-theoretic structure of combinatorial independence]
The combinatorially independent permutation criteria correspond to
the right cosets $S_{2r}/\tcal$ of $\tcal$ in $S_{2r}$. The number of
different cosets is
\begin{equation}
\frac{1}{2}{2r \choose r}\,.
\end{equation}
(The trivial criteria correspond to the coset defined by $\tcal$.)
\end{lemma}
{\bf Proof}. Let $\sigma$ and $\tau$ be two permutations that are
combinatorially dependent, i.e., there is a norm-preserving map
$\Lambda$ such that $\Lambda_\sigma(\rho)=\Lambda(\Lambda_\tau(\rho))$
for all operators $\rho$. This is equivalent to $\Lambda =
\Lambda_{\tau^{-1}\sigma}$ and consequently $\tau^{-1}\sigma$ must be
a norm-preserving permutation.

Therefore, for any given permutation $\sigma$, all permutations
belonging to the right coset defined by $\sigma$, i.e., all
permutation in the set $\{\sigma\tau\, : \, \tau\in\tcal\}$ are
combinatorially dependent. This shows that the number of
combinatorially independent criteria is equal to the number of
different right cosets $S_{2r}/\tcal$. It is given by
\begin{equation}
 |S_{2r}/\tcal|=
{|S_{2r}|\over |\tcal|}= {(2r)!\over 2 r!  r!} = {1\over 2} {2r
\choose r} \,.
\end{equation}
This ends the proof. \quad\blacksquare

One of the cosets corresponds to the group $\tcal$ of norm-preserving
permutations, so that we get ${1\over 2} {2r \choose r}-1$ non-trivial
combinatorially independent permutation separability
criteria. 

\begin{figure*}
\begin{tabular}{c|c|c}
\quad\quad graphical representation\quad\quad &
\quad\quad corresponding permutation\quad\quad & 
\quad\quad name \quad\quad \\ \hline\hline
$k\quad \bullet\longrightarrow\bullet\quad l$ & $(2k,\, 2l-1)$ & reshuffle 
$R_{kl}$ \\\hline
$k\quad\bullet\longleftarrow \bullet\quad l$ & $(2k-1,\, 2l)$ & reshuffle 
$R'_{lk}$ \\\hline
$k\quad\lar$\quad\quad\quad\quad\quad        & $(2k-1,\, 2k)$ &  \quad 
partial transpose\quad\\\hline
$k\quad\bullet$\quad\quad\quad\quad\quad     & ()          & identity
\end{tabular}
\caption{Basic permutations}
\label{tab:basicPermutations}
\end{figure*}

\begin{figure*}
\begin{tabular}{c|c}
\quad\quad valid configurations\quad\quad\quad & 
\quad\quad\quad invalid configurations\quad\quad
\\ \hline\hline
$\bullet\longrightarrow \bullet\longrightarrow \bullet$ &
$\quad \bullet\longleftarrow \bullet\longrightarrow \bullet$ 
\\ \hline
$\bullet\longleftarrow \bullet\longleftarrow\bullet$ &
$\quad \bullet\longrightarrow \bullet\longleftarrow\bullet$
\\ \hline
$\lar \quad\quad\, \bullet\longrightarrow\bullet$ & 
$\quad\lar\longleftarrow \bullet \quad\quad \bullet$
\\ \hline
$\lar \quad\quad\lar \quad\quad\bullet$ & 
$\quad\lar\longrightarrow \bullet\longrightarrow \bullet$
\end{tabular}\caption{Examples for some valid and invalid configurations of 
arrows for three subsystems}
\label{tab:configurations}
\end{figure*}

\section{Simple representatives of combinatorially independent criteria} 
Of course, one could use a computer algebra system to determine the
combinatorially independent criteria. However in such a way, one
would obtain some more or less arbitrary representatives for the right
cosets $S_{2r}/\tcal$, which do not give any insight into the
structure of the permutation criteria.  Therefore, we determine
especially simple representatives for the right cosets. To this end,
we develop a graphical notation for involutions, i.e., permutations
that can be written as a product of disjoint transpositions. The
notation is very helpful in deriving the representatives. The
graphical notation is presented in
fig.~\ref{tab:basicPermutations}. The basic permutations are identity,
partial transpose, and two types of reshuffling
\cite{ZyczkowskiB04-duality}.

For an arrow $[\,k\quad\bullet\longrightarrow\bullet\quad l\,]$ we say
that $k$ is its {\em tail} and $l$ is its {\em head}. We call an arrow
of the form $[\,k\quad \lar\,]$ a {\em loop}. For such a loop we say
that $k$ is both its head and its tail. The {\em support} of an arrow
is the set containing its head and tail. Let $C$ be a configuration of
arrows. We say that $C$ is {\em disjoint} if the supports of all pairs
of arrows are disjoint. Intuitively, this means that the arrows do not
touch each other.

\begin{theorem}[Representation by disjoint configurations]\label{thm:strzalki}
The right cosets $S_{2r}/\tcal$ can be always represented by disjoint
configurations of arrows.
\end{theorem}

To prove this theorem we will apply three basic transformations on
permutations called {\tt pruning}, {\tt chopping} and {\tt exchanging
heads}. These transformations can be realized by multiplying the
permutations with elements of the group $\tcal$ from the right. The
application of these three transformations will bring every
permutation into its normal form, i.e., an equivalent permutation that
can be represented by disjoint configurations of arrows. As an
intermediary step we will need to work with configurations of arrows
that are not necessarily disjoint. It is important that some
non-disjoint configurations of arrows should not be allowed because
they do not determine permutations unambiguously. Two arrows that
point to the same subsystem do not define a permutation unambiguously
because their corresponding transpositions are not disjoint, i.e.,
they do not commute and it would be unclear which transposition should
be applied first. Similarly, two arrows starting from the same
subsystem or an arrow starting from or pointing toward a loop are not
unambiguous. For these reasons, these configurations are not
allowed. Some of these situations are shown in
fig.~\ref{tab:configurations}. In the following 
all configurations we will use will be valid.

\begin{our_rule}[Pruning]
Let $\sigma\in S_{2r}$ be any permutation. Then we can prune all
cycles of $\sigma$ such that there are no adjacent even or odd
numbers. The pruning can be achieved by multiplication by a
norm-preserving permutation from the right.
\end{our_rule}
{\bf Proof}. Write $\sigma$ as a product of disjoint cycles.  Consider
any cycle, with two adjacent numbers $n_1$ and $n_2$: 
\begin{equation}
(k_1,\,\ldots,\, k_s,\, n_1,\, n_2,\, l_1,\, \ldots,\, l_{s'}) 
\end{equation}
We note that by applying the permutation $(n_1,\, n_2)$ from the right we
get:
\begin{eqnarray} 
&&
(k_1,\,\ldots,\, k_s,\, n_1,\, n_2,\, l_1,\, \ldots,\, l_{s'})(n_1,\, n_2) 
\nonumber \\
& = & 
(k_1,\,\ldots,\, k_s,\, n_2,\, l_1,\, \ldots,\, l_{s'})(n_1) 
\end{eqnarray}
Now, if $n_1$ and $n_2$ are either both odd or both even, then the
permutation $(n_1,\, n_2)$ is a norm-preserving permutation. Thus the
initial cycle is equivalent to the cycle with $n_1$ removed. Note that
the permutation does not affect the other cycles because the cycles
are disjoint. Therefore, the pruning process can be applied
independently to the cycles.\quad\blacksquare


\begin{figure*}[t]
\vskip0.5cm
\includegraphics[scale=0.4]{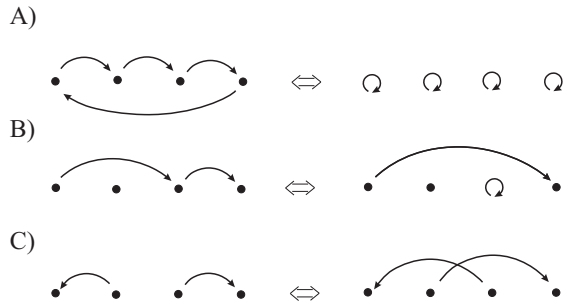}
\caption[system]{Exchanging heads.}
\label{fig:heads-change}
\end{figure*}

\begin{figure*}[t]
\vskip0.5cm
\includegraphics[scale=0.5]{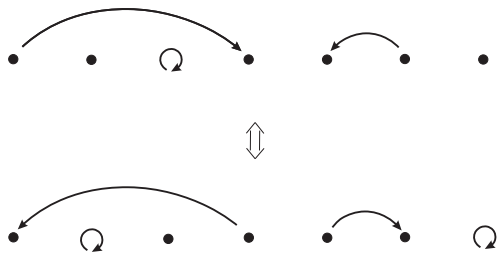}
\caption[system]{Example for the change under \flip.}
\label{fig:gt-change}
\end{figure*}

\begin{our_rule}[Chopping]
Let $\sigma\in S_{2r}$ be any permutation that cannot be further
simplified with the help of the pruning rule. Then we can chop
$\sigma$ into a product of disjoint transpositions, i.e., a
permutation that can be represented by a valid configuration of
arrows.
\end{our_rule}
{\bf Proof}. Let 
\begin{equation}
(n_1,\, p_1,\, n_2,\, p_2,\, n_3,\, p_3,\, \ldots,\, n_k,\, p_k)
\end{equation} 
be any cycle of $\sigma$. If we apply the permutation
$(n_2,\, n_1)(n_3,\, n_2) \ldots (n_1,\, n_k)$ to the above cycle then we
obtain: 
\begin{eqnarray}
&&
(n_1,\, p_1,\, \ldots,\, n_k,\, p_k)
(n_2,\, n_1)(n_3,\, n_2) \ldots (n_1,\, n_k) \nonumber \\
& = & 
(n_1,\, p_1)(n_2,\, p_2) \ldots (n_k,\, p_k) 
\end{eqnarray}
Now, if the $n_i$'s are either all even or all odd, then the applied
permutation $(n_2,\, n_1)(n_3,\, n_2) \ldots (n_1,\, n_k)$ is
norm-preserving.

After applying the chopping rule we end up with a permutation that can
be written as a product of disjoint transpositions. Therefore, the
resulting permutation can represented as a valid configuration of
arrows. \quad\blacksquare

\begin{our_rule}[Exchanging heads]
Let $C$ be a configuration of arrows. Then by exchanging the heads of
the arrows we obtain an equivalent disjoint configuration of arrows.
\end{our_rule}
{\bf Proof}. For arrows defined by $(2k,\, 2l-1)(2m,\, 2n-1)$ we take
the norm-preserving permutation $(2l-1,\, 2n-1)(2k,\, 2m)$. It is
easily verified that the operation produces arrows with exchanged
heads. Note that this works if one or two of the arrows are
loops. Some examples are presented in
fig.~\ref{fig:heads-change}. \quad\blacksquare

Before proving Theorem~1 we will provide yet another rule, 
which is not needed in the proof of theorem, but will be needed later. 
We say that the subsystem $k$ is free in a valid 
disjoint configuration if it is not contained in the support of any
arrow of $C$. In other words, a subsystem is free, when it is neither
head nor tail of any arrow (including a loop).
 
\begin{our_rule}[Flipping]
\label{rule4}
Let $C$ be a disjoint configuration. If we apply first the global quantum
transposition and then the norm-preserving permutation 
\[
\prod_{i=1}^a (2t_i-1,\, 2h_i-1)(2t_i,2h_i)\,,
\]
where $t_1\rightarrow h_1,\ldots,t_a\rightarrow h_a$ are all arrows of
$C$ that are not loops (i.e., $t_i\neq h_i$), then $C$ is changed as
follows: the directions of all arrows that are not loops are
reversed, all loops are removed, and new loops are created on all free
subsystems. 
\end{our_rule}
{\bf Proof}. Let $k$ be a free subsystem of $C$. Then we obtain
$(2k-1,\,2k)$ on subsystem $k$ by multiplying with GQT. Similarly, if
a subsystem $l$ has a loop, then the corresponding permutation
$(2l-1,2l)$ is removed by multiplying with GQT.

Let $t\rightarrow h$ be an arrow of $C$ with $t\neq h$. Then the
corresponding permutation $(2t,\,2h-1)$ is changed by multiplying with
GQT as follows
\begin{eqnarray*}
& & 
(2t,\,2h-1)(2t-1,\,2t)(2h-1,\,2h) \\
&=&
(2t,\,2h,\,2h-1,\,2t-1)\,. 
\end{eqnarray*}
Observe that there is a pair of adjacent even numbers and a pair of
adjacent odd numbers. Therefore, we can apply the pruning rule. This
is realized by multiplying with $(2t-1,\,2h-1)(2t,\,2h)$. We obtain
\begin{eqnarray*}
& & (2t,\,2h,\,2h-1,\,2t-1)(2t-1,\,2h-1)(2t,\,2h) \\
&=&
(2h,\,2t-1)\,.
\end{eqnarray*}
The resulting permutation $(2h,\,2t-1)$ is represented by the arrow
$t\leftarrow h$. This completes the proof. \quad\blacksquare

\medskip
The action of \flip\ is shown for a disjoint arrow configuration in
fig.~\ref{fig:gt-change}.  

Now, let us prove Theorem~\ref{thm:strzalki}.

\medskip
\noindent{\bf Proof of Theorem~\ref{thm:strzalki}}: Consider an
arbitrary permutation. Perform first the following steps: (1)
decompose the permutation into disjoint cycles, (2) apply pruning, and
(3) apply chopping.

After applying these three rules we have transformed the permutation
into an equivalent one that can be represented by a (valid)
configurations of arrows. Observe that configurations of arrows
consists of three types of separate objects: closed paths of arrows,
open paths of arrows, loops, and free subsystems. Now we apply the
rule exchanging heads to paths of arrows. By doing this suitably all
closed paths can be changed into a collection of loops (see
fig.~\ref{fig:heads-change}.~A), and all open paths into loops plus
one arrow - the one going from the end to the start of the open path
(see fig.~\ref{fig:heads-change}.~B). In this way, collections of all
paths are changed into sets of disjoint arrows or loops. This
completes the proof. \quad\blacksquare

\begin{table}
\begin{center}
\begin{tabular}{|c|c|}
\hline
Permutation & \\[2mm]
\hline
&\\[-3mm]
(3,12,1,2,10,8)\ (4,5,6) &  \\[2mm]
\ pruning  &  \\
$\downarrow$  &  \\[1mm]
\hline
&\\[-3mm]
(3,12,1,8)\ (5,4) &  \\[2mm]
chopping &  \\
$\downarrow$  &  \\[1mm]
\hline
&\\[-3mm]
(3,12)\ (1,8)\ (5,4) & valid configuration of arrows\\[2mm]
\ exchanging heads& \multirow{2}{*}{\ \ \includegraphics[scale=0.5]{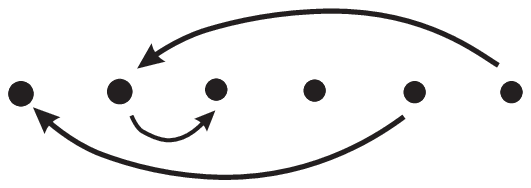} }\\
& \\[-4mm]
$\downarrow$  &  \\[2mm]
\hline
&\\[-3mm]
\multirow{2}{*}
{$\bea{ccc}
(3,4)&(1,8)&(5,12)\\
QT_2 & R_{1,5} & R_{3,6}\\
\eea$} & disjoint configuration of arrows \\
& \ \ \includegraphics[scale=0.5]{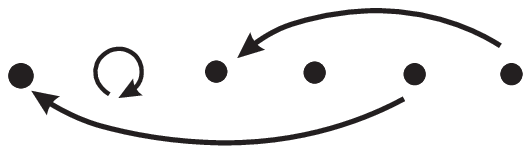} \\[3mm]
\hline
\end{tabular}
\caption[algorytm]{Steps of our algorithm in Theorem~\ref{thm:strzalki}}
\label{tab:algorythm}
\end{center}
\end{table}

\medskip
An example illustrating the steps of our algorithm for transforming
permutations to their corresponding disjoint arrow configurations in
shown in table~\ref{tab:algorythm}.

Note that after carrying out the transformation described above we can
(optionally) apply the \flip\ operation, if there are too many loops. But more
interestingly, the application of \flip, though not needed,
makes the representation much nicer. Namely, one sees that the 
transformations of exchanging heads used in the proof 
of Theorem~1 plus \flip\ act on any valid configuration (hence not necessarily 
the disjoint one) as follows:

\bee
\item every closed path of arrows disappears
\item every open path of arrows is replaced by its closing arrow
(i.e., the arrow that was missing to make the path closed)
\item every loop disappears
\item every free subsystem acquires a loop 
\eee 
Of course in the case of disjoint configurations, the closed paths do
not occur, so that the item 1 will never be used; also the item 2 will
mean changing direction of an arrow.

As a corollary of Theorem~\ref{thm:strzalki} we obtain that the group
generated in eq.~(\ref{eq:generation}) is indeed the group of
norm-preserving permutations (this is the second part of
Lemma~\ref{lem:group} whose proof was postponed).

\begin{corollary}
The group $\tcal$ of norm-preserving permutations is equal to the
group $\tcal'$ generated by
\[
(2k,\,2l),\, (2k-1,\,2l-1),\, (1,\,2)(3,\,4) \cdots (2r-1,\,2r)
\,,
\]
where $1\leq k<l\leq r$.
\end{corollary}
{\bf Proof}.  We already know that $\tcal'\subset \tcal$.
Theorem~\ref{thm:strzalki} shows that all permutations can be
represented by disjoint configurations. Obviously, all permutations in
$\tcal'$ can be represented by the empty configuration (i.e. the one
representing the trivial permutation -- the identity). Therefore, a
permutation $\sigma\not\in\tcal'$ cannot be represented by the empty
configuration. Without loss of generality $\sigma$ can be represented
by a disjoint configuration $C$ such that (1) $C$ contains an arrow
from $k$ to $l$ or (2) $C$ contains a loop on $k$ and a free subsystem
$l$ for some $k\neq l$. Now, known results for the bipartite case
imply that there are entangled states detectable by the arrow
(corresponding to realignment) in (1) and the loop and the free
subsystem (corresponding to the partial transpose) in (2). This shows
that all permutations not contained in $\tcal'$ change the norm of
some quantum states. Therefore, $\tcal'$ is equal to
$\tcal$. \quad\blacksquare

\begin{figure}[t]
\vskip0.5cm
\includegraphics[scale=0.5]{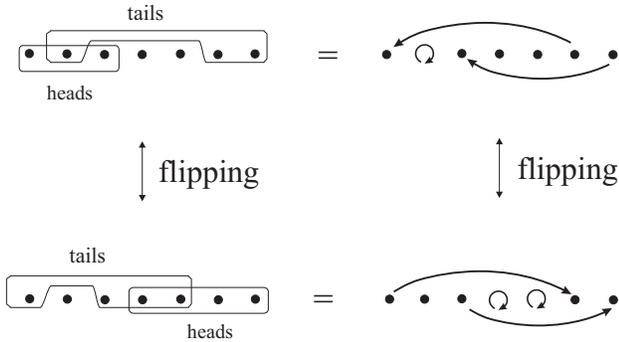}
\caption[system]{Head and tail sets of two criteria related by the rule \flip.}
\label{fig:equiv}
\end{figure}

\medskip
Let us introduce the following notation. Let $H$ and $T$ be the head
and tail sets of a disjoint configuration $C$. Denote by $\bar{H}$ and
$\bar{T}$ the head and tail sets of the configuration $\bar{C}$ that
is obtained from $C$ by applying the global quantum
transposition. Fig.~\ref{fig:equiv} shows an example of how the head
and tail sets are changed by \flip. We write $(H,T)\neq (H',T')$ to
denote that $H\neq H'$ or $T\neq T'$.

\begin{theorem}[Characterization of combinatorially independent permutation separability criteria]\label{thm:character}
All combinatorially independent separability criteria (or
equivalently all right cosets $S_{2r}/\tcal$) can be represented by a
collection of head sets $H_i$ and tail sets $T_i$ satisfying the
properties:
\begin{enumerate}
\item $(H_k,T_k)\neq (H_l,T_l)$ 
\item $(H_k,T_k)\neq (\bar{H}_l,\bar{T}_l)$
\end{enumerate}
for all $k\neq l$.
\end{theorem}

Thus, to choose combinatorially independent criteria, we first
consider all possible choices of tails and heads (of course, in each
choice, the number of heads must be equal to the number of tails).
Then we are almost done: the criteria we have obtained are pairwise
equivalent, i.e., to each criterion there is exactly one other
equivalent criterion - the one related by \flip.  Now, to get a set of
independent criteria, we keep only one criterion from each pair.

{\bf Proof of Theorem \ref{thm:character}}. Due to
Theorem~\ref{thm:strzalki} we know that all combinatorially
independent criteria can be represented by disjoint
configurations. The first condition is necessary because all
configurations with equal head and tail sets can be transformed into
each other using the rule exchanging heads. The second condition is
necessary because then configurations can be obtained by \flip.

Now we count the number of pairs of head and tail sets satisfying the
above conditions and show that it is equal to the number
$|S_{2r}/\tcal|$ of right cosets $S_{2r}/\tcal$. Therefore, the above
conditions fully characterize the combinatorially independent
criteria.

The number of pairs of head and tail sets inequivalent with respect to
the first condition is
\begin{equation}\label{eq:aleduzo}
\sum_{k=0}^r {r \choose k}{r \choose k} = {2r \choose r}\,.
\end{equation}
The summation index $k$ corresponds to the number of arrows in the
configuration. The first factorial is the number of possibilities of
choosing $k$ heads and the second of choosing $k$ tails. The equality
in (\ref{eq:aleduzo}) is a standard identity of binomial
coefficients. If we allow the second rule then must divide the
binomial coefficient ${2r\choose r}$ by $2$. But this is the number of
different cosets $S_{2r}/\tcal$. \quad\blacksquare


\def\bp{\begin{picture}} 
\def\ep{\end{picture}} 
\def\ci{\circle}
\def\li{\line} 

\begin{figure*}
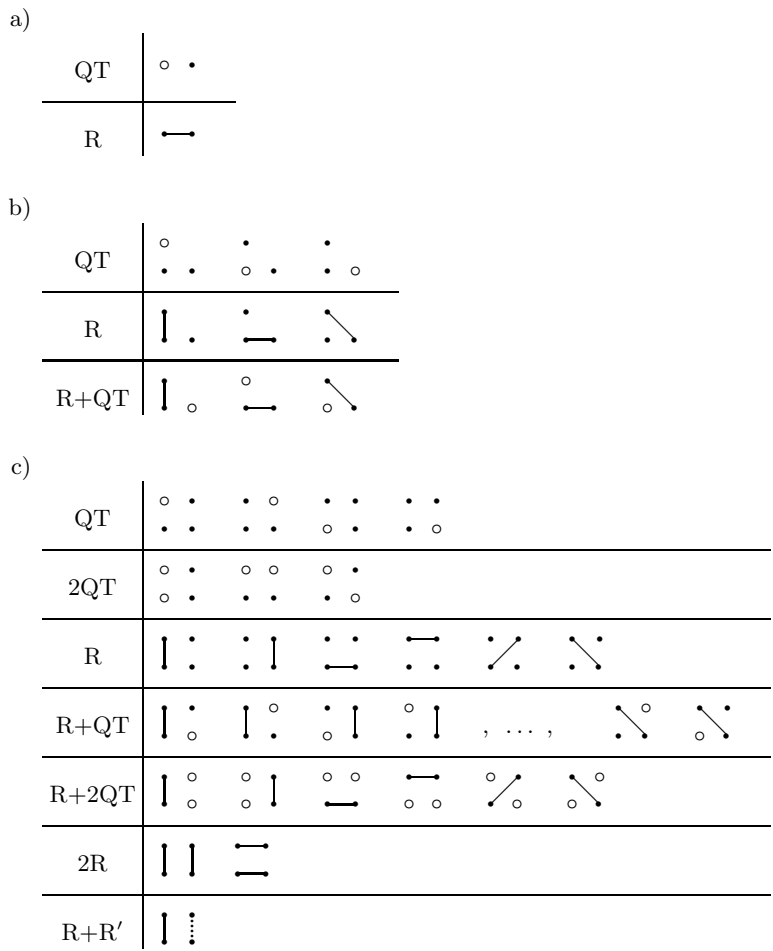

\begin{tabular}{rl}
a) & \\ &
\begin{tabular}{c|l}
\unitlength=1pt
&\\[-2mm]
\begin{minipage}{1.2cm} QT \end{minipage} & \
\ $\bp(0,0)(0,-5) \put(0,0){\ci{3}}\put(10.5,0){\ci*{2}}\ep $
\quad\quad\ \  
\\[2mm]
\hline
&\\[-2mm]
R & \
\ $\bp(0,0)(0,-5) \put(0,0){\li(1,0){10.1}}
\put(0,0){\ci*{2}}\put(10.5,0){\ci*{2}}\ep $
\quad\quad\  \  
\end{tabular}

\\
\\
b) & \\ &
\begin{tabular}{c|l}
\unitlength=1pt
&\\[-2mm]
\begin{minipage}{1.2cm} QT \end{minipage} & \
\ $\bp(0,0)(0,1) \put(0,10.5){\ci{3}}\put(0,0){\ci*{2}}\put(10.5,0){\ci*{2}}\ep $
\quad\quad\ \  
\ $\bp(0,0)(0,1) \put(0,10.5){\ci*{2}}\put(0,0){\ci{3}}\put(10.5,0){\ci*{2}}\ep $
\quad\quad\ \  
\ $\bp(0,0)(0,1) \put(0,10.5){\ci*{2}}\put(0,0){\ci*{2}}\put(10.5,0){\ci{3}}\ep $
\quad\quad\ \  
\\[2mm]
\hline
&\\[-2mm]
R & \
\ $\bp(0,0)(0,1) \put(0,0){\li(0,1){10.1}}
\put(0,10.5){\ci*{2}}\put(0,0){\ci*{2}}\put(10.5,0){\ci*{2}}\ep $
\quad\quad\  \  
\ $\bp(0,0)(0,1) \put(0,0){\li(1,0){10.1}}
\put(0,10.5){\ci*{2}}\put(0,0){\ci*{2}}\put(10.5,0){\ci*{2}}\ep $
\quad\quad\  \  
\ $\bp(0,0)(0,1) \put(0,10.5){\li(1,-1){10}}
\put(0,10.5){\ci*{2}}\put(0,0){\ci*{2}}\put(10,0){\ci*{2}}\ep $
\quad\quad\  \  
\\[2mm]
\hline
&\\[-2mm]
R+QT & \
\ $\bp(0,0)(0,1) \put(0,0){\li(0,1){10.1}}
\put(0,10.5){\ci*{2}}\put(0,0){\ci*{2}}\put(10.5,0){\ci{3}}\ep $
\quad\quad\  \  
\ $\bp(0,0)(0,1) \put(0,0){\li(1,0){10.1}}
\put(0,10.5){\ci{3}}\put(0,0){\ci*{2}}\put(10.5,0){\ci*{2}}\ep $
\quad\quad\  \  
\ $\bp(0,0)(0,1) \put(0,10.5){\li(1,-1){10}}
\put(0,10.5){\ci*{2}}\put(0,0){\ci{3}}\put(10,0){\ci*{2}}\ep $
\quad\quad\  \  

\end{tabular}

\\
\\
c) & \\ &
\begin{tabular}{c|l}
\unitlength=1pt
&\\[-2mm]
\begin{minipage}{1.2cm} QT \end{minipage} & \
\ 
$
\bp(0,0)(0,1) 
\put(0,10.5){\ci{3}}
\put(10.5,10.5){\ci*{2}}
\put(0,0){\ci*{2}}
\put(10.5,0){\ci*{2}}\ep$
\quad\quad\ \  
\ 
$
\bp(0,0)(0,1) 
\put(0,10.5){\ci*{2}}
\put(10.5,10.5){\ci{3}}
\put(0,0){\ci*{2}}
\put(10.5,0){\ci*{2}}\ep
$
\quad\quad\ \  
\ 
$
\bp(0,0)(0,1) 
\put(0,10.5){\ci*{2}}
\put(10.5,10.5){\ci*{2}}
\put(0,0){\ci{3}}
\put(10.5,0){\ci*{2}}\ep 
$
\quad\quad\ \  
\ 
$\bp(0,0)(0,1) 
\put(0,10.5){\ci*{2}}
\put(10.5,10.5){\ci*{2}}
\put(0,0){\ci*{2}}
\put(10.5,0){\ci{3}}\ep 
$
\quad\quad\ \  
\\[2mm]
\hline
&\\[-2mm]
2QT & \
\ 
$
\bp(0,0)(0,1) 
\put(0,10.5){\ci{3}}
\put(10.5,10.5){\ci*{2}}
\put(0,0){\ci{3}}
\put(10.5,0){\ci*{2}}\ep 
$
\quad\quad\  \  
\ 
$\bp(0,0)(0,1) 
\put(0,10.5){\ci{3}}
\put(10.5,10.5){\ci{3}}
\put(0,0){\ci*{2}}
\put(10.5,0){\ci*{2}}\ep 
$
\quad\quad\ \  
\ 
$\bp(0,0)(0,1) 
\put(0,10.5){\ci{3}}
\put(10.5,10.5){\ci*{2}}
\put(0,0){\ci*{2}}
\put(10.5,0){\ci{3}}\ep 
$
\quad\quad\ \  
\\[2mm]
\hline
&\\[-2mm]
R & \
\ 
$\bp(0,0)(0,1) 
\put(0,0){\li(0,1){10.1}}
\put(0,10.5){\ci*{2}}
\put(10.5,10.5){\ci*{2}}
\put(0,0){\ci*{2}}
\put(10.5,0){\ci*{2}}\ep 
$
\quad\quad\  \  
\ 
$\bp(0,0)(0,1) 
\put(10.5,0){\li(0,1){10.1}}
\put(0,10.5){\ci*{2}}
\put(10.5,10.5){\ci*{2}}
\put(0,0){\ci*{2}}
\put(10.5,0){\ci*{2}}\ep 
$
\quad\quad\  \  
\ 
$\bp(0,0)(0,1) 
\put(0,0){\li(1,0){10.1}}
\put(0,10.5){\ci*{2}}
\put(10.5,10.5){\ci*{2}}
\put(0,0){\ci*{2}}
\put(10.5,0){\ci*{2}}\ep 
$
\quad\quad\  \  
\ 
$\bp(0,0)(0,1) 
\put(0,10.5){\li(1,0){10.1}}
\put(0,10.5){\ci*{2}}
\put(10.5,10.5){\ci*{2}}
\put(0,0){\ci*{2}}
\put(10.5,0){\ci*{2}}\ep 
$
\quad\quad\  \  
\ 
$\bp(0,0)(0,1) 
\put(0,0){\li(1,1){10}}
\put(0,10.5){\ci*{2}}
\put(10.5,10.5){\ci*{2}}
\put(0,0){\ci*{2}}
\put(10,0){\ci*{2}}\ep 
$
\quad\quad\  \  
\ 
$\bp(0,0)(0,1) 
\put(0,10.5){\li(1,-1){10}}
\put(0,10.5){\ci*{2}}
\put(10.5,10.5){\ci*{2}}
\put(0,0){\ci*{2}}
\put(10,0){\ci*{2}}\ep $
\quad\quad\  \  
\\[2mm]
\hline
&\\[-2mm]
R+QT &  \
\ $\bp(0,0)(0,1) \put(0,0){\li(0,1){10.1}}
\put(0,10.5){\ci*{2}}\put(10.5,10.5){\ci*{2}}\put(0,0){\ci*{2}}\put(10.5,0){\ci{3}}\ep $
\quad\quad\  \  
\ $\bp(0,0)(0,1) \put(0,0){\li(0,1){10.1}}
\put(0,10.5){\ci*{2}}\put(10.5,10.5){\ci{3}}\put(0,0){\ci*{2}}\put(10.5,0){\ci*{2}}\ep $
\quad\quad\  \  
\ $\bp(0,0)(0,1) \put(10.5,0){\li(0,1){10.1}}
\put(0,10.5){\ci*{2}}\put(10.5,10.5){\ci*{2}}\put(0,0){\ci{3}}\put(10.5,0){\ci*{2}}\ep $
\quad\quad\  \  
\ $\bp(0,0)(0,1) \put(10.5,0){\li(0,1){10.1}}
\put(0,10.5){\ci{3}}\put(10.5,10.5){\ci*{2}}\put(0,0){\ci*{2}}\put(10.5,0){\ci*{2}}\ep $
\quad\quad\  \  
, \  \ldots \ ,  \quad\quad
\ $\bp(0,0)(0,1) \put(0,10.5){\li(1,-1){10}}\put(0,10.5){\ci*{2}}\put(10.5,10.5){\ci{3}}\put(0,0){\ci*{2}}\put(10,0){\ci*{2}}\ep $
\quad\quad\  \  
\ $\bp(0,0)(0,1) \put(0,10.5){\li(1,-1){10}}\put(0,10.5){\ci*{2}}\put(10.5,10.5){\ci*{2}}\put(0,0){\ci{3}}\put(10,0){\ci*{2}}\ep $
\quad\quad\  \  
\\[2mm]
\hline
&\\[-2mm]
R+2QT & \
\ $\bp(0,0)(0,1) \put(0,0){\li(0,1){10.1}}
\put(0,10.5){\ci*{2}}\put(10.5,10.5){\ci{3}}\put(0,0){\ci*{2}}\put(10.5,0){\ci{3}}\ep $
\quad\quad\  \  
\ $\bp(0,0)(0,1) \put(10.5,0){\li(0,1){10.1}}
\put(0,10.5){\ci{3}}\put(10.5,10.5){\ci*{2}}\put(0,0){\ci{3}}\put(10.5,0){\ci*{2}}\ep $
\quad\quad\  \  
\ $\bp(0,0)(0,1) \put(0,0){\li(1,0){10.1}}
\put(0,10.5){\ci{3}}\put(10.5,10.5){\ci{3}}\put(0,0){\ci*{2}}\put(10.5,0){\ci*{2}}\ep $
\quad\quad\  \  
\ $\bp(0,0)(0,1) \put(0,10.5){\li(1,0){10.1}}
\put(0,10.5){\ci*{2}}\put(10.5,10.5){\ci*{2}}\put(0,0){\ci{3}}\put(10.5,0){\ci{3}}\ep $
\quad\quad\  \  
\ $\bp(0,0)(0,1) \put(0,0){\li(1,1){10}}
\put(0,10.5){\ci{3}}\put(10.5,10.5){\ci*{2}}\put(0,0){\ci*{2}}\put(10,0){\ci{3}}\ep $
\quad\quad\  \  
\ $\bp(0,0)(0,1) \put(0,10.5){\li(1,-1){10}}
\put(0,10.5){\ci*{2}}\put(10.5,10.5){\ci{3}}\put(0,0){\ci{3}}\put(10,0){\ci*{2}}\ep $
\quad\quad\  \  
\\[2mm]
\hline
&\\[-2mm]
2R & \
\ \bp(0,0)(0,1) \put(0,0){\li(0,1){10.1}}\put(10.5,0){\li(0,1){10.1}}
\put(0,10.5){\ci*{2}}\put(10.5,10.5){\ci*{2}}\put(0,0){\ci*{2}}\put(10.5,0){\ci*{2}}\ep 
\quad\quad\  \  
\ \bp(0,0)(0,1) \put(0,0){\li(1,0){10.1}}\put(0,10.5){\li(1,0){10.1}}
\put(0,10.5){\ci*{2}}\put(10.5,10.5){\ci*{2}}\put(0,0){\ci*{2}}\put(10.5,0){\ci*{2}}\ep 
\quad\quad\  \  
\\[2mm]
\hline
&\\[-2mm]
R+R${}'$ & \ 
\ $\bp(0,0)(0,1) \put(0,0){\li(0,1){10.1}}
\multiput(10.5,0.25)(0,2){6}{\ci*{0.5}}
\put(0,10.5){\ci*{2}}\put(10.5,10.5){\ci*{2}}\put(0,0){\ci*{2}}\put(10.5,0){\ci*{2}}\ep $
\quad\quad\  \  \\
\end{tabular}
\end{tabular}
\caption[four]{Combinatorially independent permutation criteria for
a) two, b) three, and c) four particles. Bars represent reshuffling
$R$ and open circles represent quantum transpositions. The dotted line
in c) represents the second type of reshuffling $R'$.}
\label{fig:four}
\end{figure*}

\section{Examples of combinatorially independent criteria}

In this section we present all combinatorially independent permutation
criteria for two, three and four particles. The arrow notation we have
used so far is convenient for singling out combinatorially independent
criteria. However, the arrow notation is based on a linear ordering of
the particles, which is not convenient for visualizing the chosen
independent criteria.
\[
\lar\equiv \bp(0,0)(0,1)\put(0,3){\circle{3}}\ep\quad,\quad
\bullet \longrightarrow \bullet \equiv \ \ {\bp(0,0)(0,1) \put(0,3){\line(1,0){10.1}}
\put(0,3){\circle*{3}}\put(10.5,3){\circle*{3}}\ep}\quad \quad,\quad
\bullet \longleftarrow \bullet \equiv \ \ \bp(0,0)(0,1) 
\put(0,3){\circle*{3}}\put(10.5,3){\circle*{3}}
\multiput(0,3)(2,0){6}{\ci*{0.5}}\ep
\]

The criteria are visualized in fig.~\ref{fig:four}. For three
particles, all criteria are combinations of reshuffling $R$ and
partial transpose. For four particles a new possibility appears, which
is compatible with Ref. \cite{Fan-permut}. We see that only one
criterion uses different reshuffling $R'$. Conversely, the application
of GQT would lead to a complementary set of criteria, where only one
reshuffling $R$ would be present.

\section{Concluding remarks}
We have analyzed and simplified the structure of the permutation
criteria. To this end, we have introduced the notion of combinatorial
independence that is necessary for two permutation to yield
independent separability criteria and completely characterized all
combinatorially independent permutation criteria. The especially
simple form of our representatives of all combinatorially independent
criteria show explicitly how the permutation criteria probe the
quantum state: to use the full power of permutation criteria, we can
restrict ourselves to simple building blocks given by partial
transpose and two types of reshuffling. This could be the starting
point toward fully understanding the power of the permutation
separability criteria, i.e., determining if there are entangled states
that cannot be detected by any permutation criterion and establishing
which permutations define truly independent separability criteria. In
particular, our results simplify the evaluation of candidates for the
entanglement measure proposed in \cite{HHH02-permut}. Perhaps our
tools could enable a qualitative classification of multipartite
entanglement with respect to some combinatorial classes.

\medskip
{\em Acknowledgments}. We would like to thank Asa Ericsson and Lieven
Clarisse for useful comments. The paper has been written while
M.~H. was visiting the Institute for Quantum Information, California
Institute of Technology. M.~H. is supported by Polish Ministry of
Scientific Research and Information Technology under the (solicited)
grant no.~PBZ-MIN-008/P03/2003 and by EC grants RESQ, contract
no.~IST-2001-37559 and QUPRODIS, contract no.~IST-2001-38877. P.W. is
supported by the National Science Foundation under the grant no.~EIA
0086038.


\begin{thebibliography}{BDH{\etalchar{+}}02}

\bibitem[ABH{\etalchar{+}}01]{Alber2001}
Gernot Alber, Thomas Beth, Micha\l{} Horodecki, Pawe\l{} Horodecki, Ryszard
  Horodecki, Martin R{\"o}tteler, Harald Weinfurter, Reinhard Werner, and Anton
  Zeilinger.
\newblock {\em Quantum Information: An Introduction to Basic Theoretical
  Concepts and Experiments}.
\newblock Springer, 2001.

\bibitem[BCH{\etalchar{+}}02]{Bruss2001-reflections}
Dagmar Bruss, J.~I. Cirac, Pawe\l{} Horodecki, F.~Hulpke, Barbara Kraus, Maciej
  Lewenstein, and Anna Sanpera.
\newblock Reflections upon separability and distillability.
\newblock {\em J. Mod. Opt.}, 49:1399--1418, 2002.
\newblock quant-ph/0110081.

\bibitem[BDH{\etalchar{+}}02]{BadziagDHHH01-conc}
Piotr Badziag, Piotr Deuar, Michal Horodecki, Pawel Horodecki, and Ryszard
  Horodecki.
\newblock Concurrence in arbitrary dimensions.
\newblock {\em J. Mod. Opt.}, 49:1289--1297, 2002.
\newblock quant-ph/0107147.

\bibitem[CW02]{ChenWu02-multisep}
Kai Chen and Ling-An Wu.
\newblock The generalized partial transposition criterion for separability of
  multipartite quantum states.
\newblock {\em Phys. Lett. A}, 306:14--20, 2002.
\newblock quant-ph/0208058.

\bibitem[CWY03]{Chen2002-criterion}
Kai Chen, Ling-An Wu, and Li~Yang.
\newblock A matrix realignment method for recognizing entanglement.
\newblock quant-ph/0205017, 2003.

\bibitem[DPS04]{DohertyPS04}
Andrew~C. Doherty, Pablo~A. Parrilo, and Federico~M. Spedalieri.
\newblock Detecting multipartite entanglement.
\newblock quant-ph/0407143, 2004.

\bibitem[Fan02]{Fan-permut}
Heng Fan.
\newblock A note on separability criteria for multipartite state.
\newblock quant-ph/0210168, 2002.

\bibitem[GL04]{GuehneL04-sep}
Otfried Guehne and Maciej Lewenstein.
\newblock Separability criteria from uncertainty relations.
\newblock {\em AIP Conf. Proc.}, 734:230, 2004.
\newblock quant-ph/0409140.

\bibitem[HH94]{HH94-redun}
Pawe\l{} Horodecki and Ryszard Horodecki.
\newblock Quantum redundancies and local realism.
\newblock {\em Phys. Lett. A}, 194:147--152, 1994.

\bibitem[HHH96]{sep1996}
Micha\l{} Horodecki, Pawe\l{} Horodecki, and Ryszard Horodecki.
\newblock Separability of mixed states: Necessary and sufficient conditions.
\newblock {\em Phys. Lett. A}, 223:1, 1996.
\newblock quant-ph/9605038.

\bibitem[HHH02]{HHH02-permut}
Michal Horodecki, Pawel Horodecki, and Ryszard Horodecki.
\newblock Separability of mixed quantum states: linear contractions approach.
\newblock quant-ph/0206008, 2002.

\bibitem[MKB04]{MintertKB04-conc}
Florian Mintert, Marek Kus, and Andreas Buchleitner.
\newblock Concurrence of mixed bipartite quantum states in arbitrary
  dimensions.
\newblock {\em Phys. Rev. Lett}, 92:167902, 2004.
\newblock quant-ph/0403063.

\bibitem[Per96]{Peres96}
Asher Peres.
\newblock Separability criterion for density matrices.
\newblock {\em Phys. Rev. Lett.}, 77:1413, 1996.

\bibitem[QIC]{QIC}
First volume of Quantum Information \& Computation, 2001.

\bibitem[Rud02]{Rudolph2002-criterion}
Olivier Rudolph.
\newblock Further results on the cross norm criterion for separability.
\newblock quant-ph/0202121, 2002.

\bibitem[Wer89]{Werner1989}
Reinhard Werner.
\newblock Quantum states with Einstein-Rosen-Podolsky correlations admitting a
  hidden-variable model.
\newblock {\em Phys. Rev.}, A 40:4277, 1989.

\bibitem[ZB04]{ZyczkowskiB04-duality}
Karol Zyczkowski and Ingemar Bengtsson.
\newblock On duality between quantum maps and quantum states.
\newblock {\em Open Sys. Inf. Dyn.}, 11:3--42, 2004.
\newblock quant-ph/0401119.

\end{thebibliography}

\newcommand{\etalchar}[1]{$^{#1}$}

\end{document}